\documentclass[aip,jcp,reprint,citeautoscript]{revtex4-1} 
\usepackage[T1]{fontenc}
\usepackage{verbatim}
\usepackage{amsmath}
\usepackage{amssymb}
\usepackage{graphicx}
\usepackage[mathscr]{euscript}
\makeatletter

\newcommand{\ket}[1]{|#1\rangle}             
\newcommand{\braket}[2]{\langle#1|#2\rangle} 

\providecommand{\figwidth}{3.2in}
\graphicspath{{Figures/}}

\makeatother

\begin{document}

\title{Practical Witness for Electronic Coherences}

\author{Allan S.~Johnson}
\affiliation{Department of Physics, University of Ottawa, Ottawa, Canada}
\affiliation{Department of Physics, Imperial College London, London, UK}
\author{Joel Yuen-Zhou}
\affiliation{Center for Excitonics, Research Laboratory of Electronics,
Massachusetts Institute of Technology, Cambridge, USA}
\author{Al\'an Aspuru-Guzik}
\affiliation{Department of Chemistry, Harvard University, Cambridge, USA}
\author{Jacob J.\ Krich}
\affiliation{Department of Physics, University of Ottawa, Ottawa, Canada}

\begin{abstract}
The origin of the coherences in two-dimensional spectroscopy of photosynthetic
complexes remains disputed. Recently it has been shown that in the
ultrashort-pulse limit, oscillations in a frequency-integrated pump-probe signal correspond
exclusively to electronic coherences, and thus such experiments can
be used to form a test for electronic vs. vibrational oscillations
in such systems. Here we demonstrate a method for practically implementing
such a test, whereby pump-probe signals are taken at several different
pulse durations and used to extrapolate to the ultrashort-pulse limit.
We present analytic and numerical results determining requirements
for pulse durations and the optimal choice of pulse central frequency,
which can be determined from an absorption spectrum. Our results suggest that for numerous systems the required experiment could be implemented by many ultrafast spectroscopy laboratories using pulses of tens of femtoseconds in duration. Such experiments could resolve the standing debate over the nature of coherences in photosynthetic complexes.
\end{abstract}

\maketitle

\section{Introduction}

While it was often assumed that coherent quantum dynamics would not
persist at physiological temperatures, the discovery of oscillatory
features in two-dimensional spectroscopy of photosynthetic systems\citep{Cina2004,Brixner2005,Scholes2006,Engel2007,Collini2010,Sarovar2010,Panitchayangkoon2011,Hayes2013,Ostroumov2013,Hildner2013}
have made such systems one of the prototypes for the nascent field
of quantum biology \citep{Panitchayangkoon2010,Lambert2012}. Of particular interest is the potential influence of quantum behavior on the highly efficient energy transport between the absorption and reaction sites in photosynthetic complexes \citep{Kassal2013,Brumer2012,Schlawin2013}. 
The impact of these coherent signatures upon energy transport has
been heavily studied, and depends greatly upon the type of coherences observed \citep{Wu2012,Mohseni2008,Plenio2008,Rebentrost2009,Gelin2012,Christensson12,Chin13,Chenu13,Tiwari13}.
For this reason, a simple method for distinguishing between electronic and vibrational coherences could play an important role in the study of these quantum biological systems. While quantum process tomography \citep{Rebentrost2011,Yuen-Zhou2011,Yuen-Zhou2011a} or wavepacket
reconstruction techniques \citep{Avisar2011,Avisar2011a,Humble2004,Biggs2009}
could provide such a method, these techniques come at the cost of performing several experiments \citep{Yuen-Zhou2014, Perlik2014}.

An alternative technique is based on pump-probe spectroscopy. Stock and Domcke showed over twenty-five years ago that, for a system without electronic coherences, the frequency-integrated pump-probe signal in the impulsive limit shows no dependence on time delay, in the Condon approximation \citep{Stock1988,Stock92}. For studies of vibrational dynamics, this effect is undesirable \citep{Jonas95}. In Ref.\ \citenum{Yuen-Zhou2012}, it was realized that this effect could be applied to the problem of distinguishing electronic and vibrational coherences. The ease of a pump-probe experiment is a significant advantage over other methods proposed for disentangling contributions from vibrational and electronic coherences \cite{Perlik2014,Halpin14}. Real experiments, however, do not have infinitely short, impulsive pulses. Here we show that, using only finite-bandwidth pulses, vibrational and electronic coherences can be discriminated with extreme prejudice. Indeed, the use of ultrashort-pulse pump-probe experiments to discriminate between
types of coherences lends itself to a simple recipe, which we outline here: (1) Use the absorption spectrum of the system to determine the pulse central frequencies and duration requirements. (2) Perform several pump-probe experiments with pulses of decreasing duration, each with a full width half maximum (FWHM) less than $\sqrt{2\ln2}/5\Sigma_{A}$, where $\Sigma^2_{A}$ is the variance of the absorption spectrum (defined in Sec.\ \ref{section:optimal}). (3) Considering only waiting (delay) times without significant pulse-overlap effects, subtract the average pump-probe signal, and use the square integral of the remaining signal  as a measure of the magnitude of oscillations. Make a ``witness plot'' of this oscillation amplitude against the duration of the pulses used. If the oscillation amplitude decreases
monotonically with decreasing pulse duration (i.e., has positive slope), the signal is vibrational in origin; an electronic or a mixture of electronic and vibrational coherences will increase monotonically. Each of these steps is elaborated upon and demonstrated in the follow discussion. We provide instructions suitable
for experimental implementation. Numerical simulations performed in model systems indicate that with properly centered pulses, pulses of sub-120fs (FWHM) in duration should be sufficient for discriminating the two types of coherences in many systems.  

\section{Background and Model}
\label{section:background}

In previous work it was shown that, within the Condon approximation
and the ultrashort-pulse limit, oscillations in the pump-probe signal
correspond exclusively to electronic oscillations. More precisely,
the pump-probe signal can be written in the reduced electronic basis
of the singly excited states \citep{Yuen-Zhou2012}. This result can
be understood from a simple physical picture. A molecular system is
initially in its (non-degenerate) ground electronic state and some
vibrational state. An ultrashort pump pulse produces a vibronic excitation in an excited electronic state. If the pump pulse is sufficiently short, the nuclear coordinates are frozen during the excitation.
The excited wavepacket then evolves in the excited state potential energy surface for some time $T$ before the probe pulse causes another transition. If the probe pulse is sufficiently short that the nuclear coordinates are frozen during this transition and the transition dipole is independent of the nuclear coordinates (Condon approximation), then the probability of wavepacket transfer is independent of nuclear coordinate, and the only interference that can occur is between different electronic states. Oscillations in the pump-probe signal are thus attributable to electronic coherences alone in this limit. 

Reference \citenum{Yuen-Zhou2012}  suggested this technique could be extended to pulses of finite duration by collecting the pump-probe signals with pulses of several different durations, extrapolating to the impulsive limit. Here we confirm the success of pulse-duration extrapolation, and discuss the optimal experimental parameters with both analytical results and numerical simulations; in particular we highlight the maximum pulse duration for which the test will function, which we call the witness time $T_W$.

We begin by describing the extrapolation procedure using numerical
data, which also illustrates the issue of the witness time. We demonstrate this procedure using numerical simulations on simple systems consisting of either a monomer (electronic two-level system) or dimer (electronic four-level system) coupled to one or two harmonic vibrational modes, respectively. Simulations were performed using wavepacket propagation techniques and calculating wavepacket overlaps; further details on numerical methods and parameters can be found in Appendix \ref{appendix:numerics} and in chapter 6 of Ref.~\citenum{Yuen-Zhou2014}. 

We assume that the energy scale separating ground and excited electronic states is much larger than the vibrational energies, and we assume the rotating wave approximation for interaction with the light. The vibrational potential energy surfaces of the excited electronic states may have different frequencies ($\omega_e$) and equilibrium vibrational coordinates ($\Delta_x$), from the ground state vibrational surface. The monomer system with a single vibrational mode can then be described by a Hamiltonian of the form
\begin{align}
H=\frac{p^2}{2}+\frac{\omega_0^2x^2}{2}|g\rangle\langle g|+\left(\Omega_e+\frac{\omega_e^2(x-\Delta_x)^2}{2}\right)|e\rangle\langle e|
\end{align}
where $x$ is the nuclear coordinate including the particle masses, $p$ is the nuclear momentum, $|i\rangle$ denotes an electronic state, $\omega_0$ is the ground state vibrational frequency, $\Omega_e$ is the electronic excitation energy, and Planck's constant $\hbar$ is set to 1. A diagram of the electronic potential surfaces of this system is shown in Figure \ref{fig:potential_surfaces}. The extension to a dimer is straightforward; see Appendix \ref{appendix:numerics} for the full dimer Hamiltonian (equation \ref{eq:dimer_hamil}). 

\begin{figure}[h]
\includegraphics[width=\linewidth]{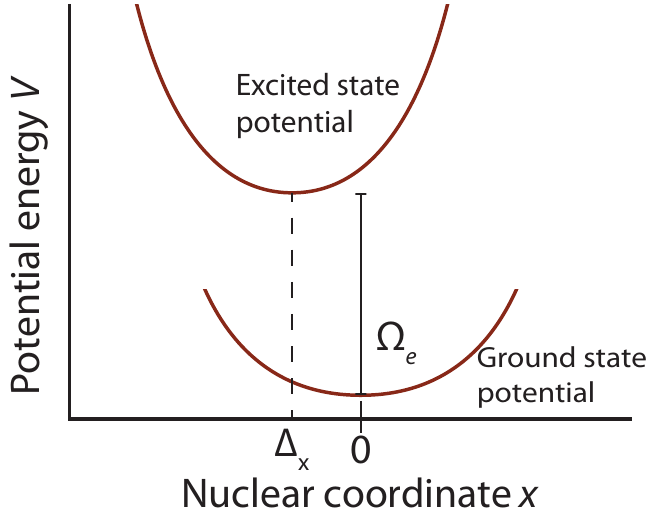}
\caption{Potential energy surfaces for a monomer with a single nuclear degree of freedom, not to scale. The electronic energy gap $\Omega_e$ is much larger than the nuclear energy scale. The equilibrium position of the excited state surface is shifted by $\Delta_x$ from the ground state equilibrium.}
\label{fig:potential_surfaces} 
\end{figure}

We treat the pump and probe pulses as classical Gaussian electric
fields with magnitude at the position of the molecules $\epsilon_{Q}(t)=\frac{\eta}{\sqrt{2\pi\sigma_{Q}^{2}}}e^{-i\omega_{Q}(t-t_{Q})}e^{-(t-t_{Q})^{2}/2\sigma_{Q}^{2}}+c.c.$, where $Q=P,P^{\prime}$ for the pump and probe pulses, respectively, with central frequencies $\omega_{Q}\approx\Omega_e$, durations $\sigma_{Q}$, pulse times $t_{P}=0$, $t_{P^{\prime}}=T$, and pulse strength $\eta$. We treat the interaction of these pulses with the system within the dipole approximation. We further assume the Condon approximation, namely, that the electronic dipole operator is independent of vibrational state, and neglect relaxation and dephasing effects other that those caused by the explicitly modeled vibrational modes. We choose units such that $\omega_0=1$, so frequencies are expressed as multiples of $\omega_0$ and times as multiples of $\omega_0^{-1}$. 
For example, in a system with a soft vibrational mode of 100~cm$^{-1}$, times are expressed in units of 
$1/\omega_0 = 1/(100\times2\pi c)$~cm$ = 53$~fs. We describe the shift in equilibrium vibrational coordinate using both $\Delta_x$ and the  dimensionless Huang-Rhys factor $\mathscr{S}=\Delta_x^2/2\omega_0$ \citep{Huang50}.

\begin{figure}[h]
\centering \includegraphics[width=\figwidth]{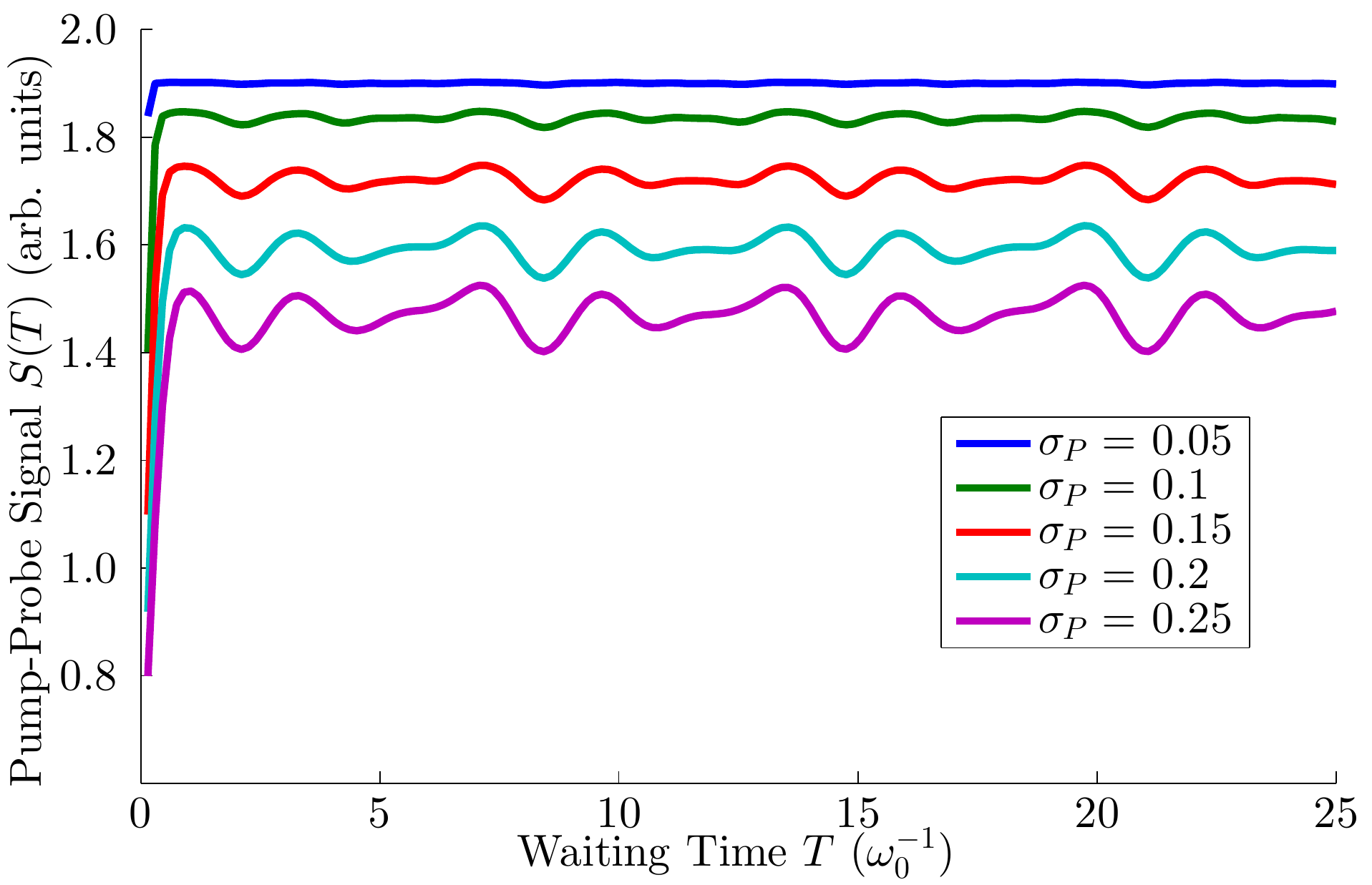}
\caption{\label{PumpProbe}Pump-probe signal for a sample monomer with equal
pump and probe durations, as given in the legend in units of $\omega_{0}^{-1}$.
At long pulse durations the spectrum exhibits oscillations, but as
the duration is decreased the oscillations' visibility decrease monotonically. Here $\omega_{e}=1.5\omega_0$ and $\mathscr{S}=0.02$. In this simulation the system begins in the fourth vibrational state; the same results hold in the ground state case, but the oscillations are less visible due to a larger change of the signal in the pulse-overlap region. For the ground state case, as well as a full thermal and isotropic simulation of the same system, see Fig.~\ref{Witness_thermal}.}
\end{figure}

Our monomer has only one electronic excited state and thus has only
vibrational coherences. It is therefore the simplest test case for
the witness, and we expect its pump-probe signal to exhibit no oscillations, for sufficiently short pulses. Figure \ref{PumpProbe} shows the numerically-determined pump-probe signal for a monomer with a harmonic vibrational mode for various pulse durations. To make the oscillatory effects more visible, we show here the case where the system is initially in the third vibrational state. In all subsequent simulations, unless otherwise noted, we begin in the ground vibrational state. For a simulation including thermal and isotropic averaging, see Fig.~\ref{Witness_thermal}. The pump-probe signal has clear oscillatory components for each pulse duration. The visibility of these oscillations decreases as the pulse duration decreases, illustrating the value of extrapolation to an ultrashort pulse.

\section{Analyzing the Experiments: Witness Plots}
We are now in a position to discuss how to create an unambiguous test for discriminating electronic and vibrational coherences through extrapolation to the impulsive limit. To begin, we note that in Figure \ref{PumpProbe}, there is a steep increase
in the pump-probe signal $S_{PP}(T,\sigma_P,\sigma_{P'})$ at short waiting times. This steep increase is due to temporal overlap of the pump and probe pulses, during which the pump pulse is steadily promoting amplitude to the excited state(s), and does not reflect the dynamics of interest. We remove these pulse-overlap effects by analyzing only data after  $T_{min}=3(\sigma_{P,max}+\sigma_{P^{\prime},max})$,
where the ``max'' indicates that the same time cutoff is used for
all choices of pulse duration. We then subtract the mean of each remaining signal, leaving only oscillatory features. The oscillatory features at this point correspond in general to both vibrational and electronic coherences, though for the monomer they are solely vibrational in origin. We obtain a measure of the oscillations by integrating over the magnitude of the oscillatory portion, giving a total signal of the following form:
\begin{align}
	\Gamma(\sigma_P,\sigma_P') = \int_{T_{min}}^{T_{final}} \left\vert S_{PP}(T,\sigma_P,\sigma_P')-\bar{S}_{PP}(\sigma_P,\sigma_P') \right\vert^2 dT,
\end{align} 
where $\bar{S}_{PP}(\sigma_P,\sigma_P')$ is the mean value of $S_{PP}$ for $T>T_{min}$. 

\begin{figure}[h]
\centering \centering \includegraphics[width=\figwidth]{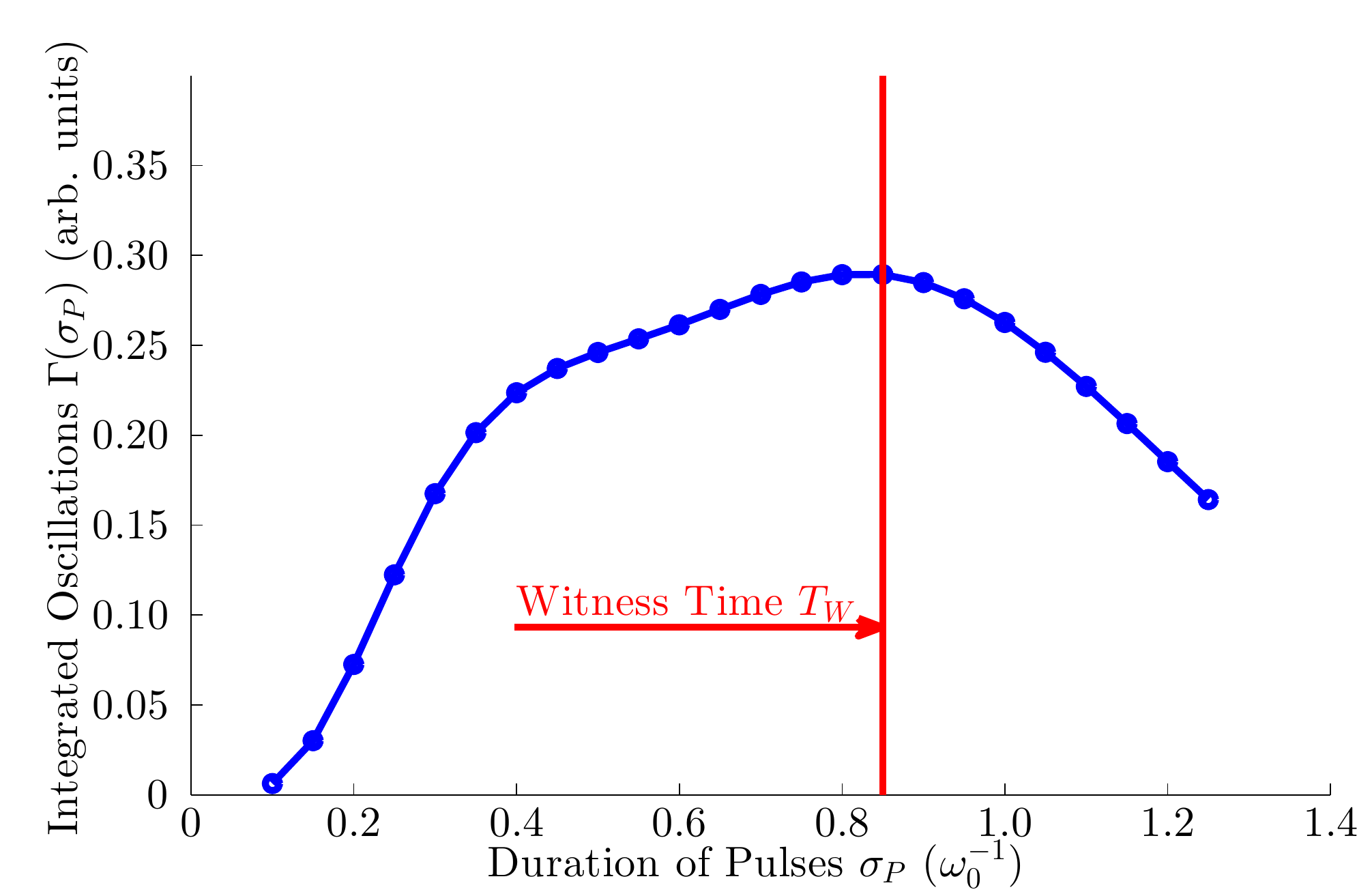}

\caption{Witness plot, i.e., magnitude of oscillations vs.~pulse duration, for a sample monomer, corresponding to the simulated data of Fig.~\ref{PumpProbe}. As the pulse duration decreases to zero, the oscillations decrease. Monotonic behavior is observed only once the the pulse duration is below the indicated witness time $T_W$.}

\label{Witness_monomer} 
\end{figure}

We plot $\Gamma(\sigma_P,\sigma_P')$ against the duration of
the pulses used with $\sigma_P=\sigma_{P'}$, as in Fig.~\ref{Witness_monomer}. If the result
has positive slope (i.e., monotonically decreases with decreasing
pulse duration), the coherences are vibrational; if the signal has
negative slope, electronic coherences are present. Note that for a
system with a mixture of vibrational and electronic coherences, this technique can be generalized by Fourier-transforming the pump-probe signal with respect to the waiting time $T$, selecting a frequency peak, and plotting its amplitude
as a function of the pulse duration used to obtain the pump-probe
signal\cite{Halpin14, Goodknight2014}. This generalized method allows for the nature of a particular frequency response to be determined. However, if we are interested only in whether the system has electronic coherences, the total signal can be used instead. We will proceed under the assumption that the question of interest is whether or not electronic coherences are present.

Figure \ref{Witness_monomer} shows the result of this process for
the monomer system with the same parameters as in Fig.~\ref{PumpProbe};
as the system in question is a monomer, the oscillatory response of
the system decreases as the pulse durations approach zero. We refer
to these plots of the oscillatory behavior as a function of pulse
duration as ``witness plots,'' as they provide the witness for the
nature of the coherence.

Figure \ref{Witness_monomer} shows that the integrated oscillatory signal has positive slope only for pulse durations less than a critical pulse duration, which we call the witness time $T_{W}$, as indicated in the figure. For longer pulse durations, oscillations in the pump-probe signal decrease in magnitude because the pulse duration approaches the vibrational period. This leads to an averaged pump-probe signal that becomes independent of $T$ in the long pulse limit; this situation gives a declining oscillatory signal as $\sigma_{P}$, $\sigma_{P'}$ increase, which is unrelated to the physics of the witness and was described previously  \citep{Stock1988,Jonas95}. In Fig.\ \ref{Witness_monomer}, $T_W$ is of the same order as $\omega_0^{-1}$, which illustrates the general result that the pulse durations must be shorter than the vibrational period $2\pi/\omega_0$.
Only experiments performed
in the witness region (i.e., with pulse durations less than the witness
time) can effectively discriminate between electronic and vibrational
coherences. The determination of the witness time is thus of vital
importance for any experimental implementation of the technique. The selection of the central frequency of the pulses can affect the
witness time; we will discuss optimal experimental parameters for
implementation of the witness in Section \ref{section:optimal}. 

\begin{figure}[h]
\centering \centering \includegraphics[width=\figwidth]{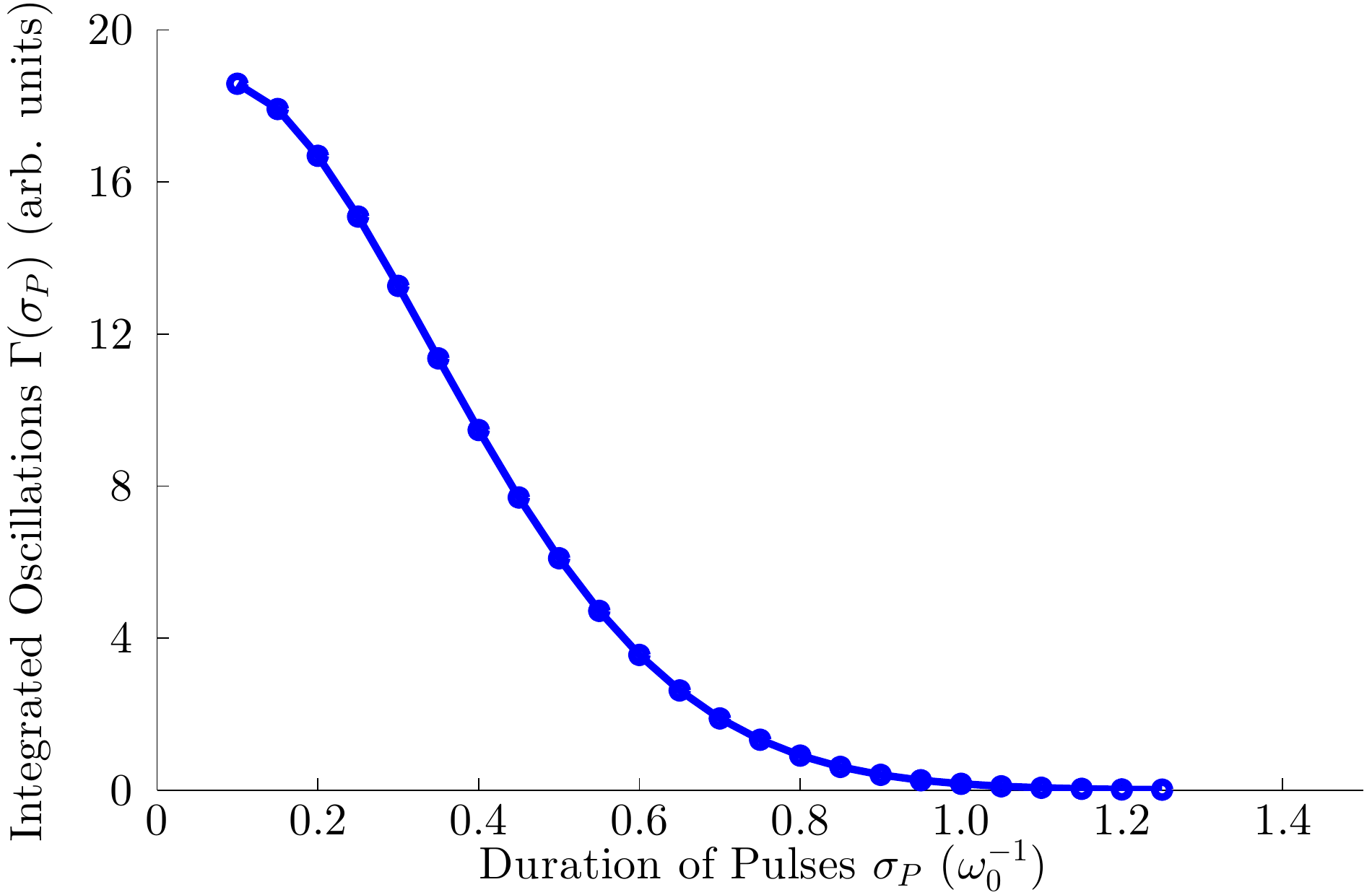}

\caption{Witness plot for a simulated dimer, with parameters $J=1 \omega_0$, $\mathscr{S}_1=0.02$, $\omega_{e,1}=1.5 \omega_0$, $\mathscr{S}_2=0.005$, $\omega_{e,2}=2 \omega_0$, and $\Delta E_{1,2}=0.73 \omega_0$ (see Appendix \ref{appendix:numerics} for details). The amplitude of oscillations in the pump-probe signal increases as the pulse duration decreases, in contrast to the case of the monomer shown in Fig.~\ref{Witness_monomer}, thus indicating a positive witness for electronic coherence.}

\label{Witness_dimer} 
\end{figure}

Figure \ref{Witness_dimer} shows a witness plot for a coupled dimer, a system with both vibrational and electronic coherences. The two
singly excited electronic states are coupled by an electronic coupling
$J$, as in {Ref.~}\citenum{Yuen-Zhou2012}. As there are two coupled
singly excited electronic states with different transition dipoles, we expect to see oscillatory components persist even with small pulse durations. 
%
%
In Figure \ref{Witness_dimer} the integrated oscillatory signal has negative slope throughout, in contrast to the monomer case, thus demonstrating a positive witness for electronic coherence. In a wide variety of prototypical systems simulated we have found this qualitative behaviour in all those with electronic coherences. This qualitatively different behavior in the witness plots thus provides a method to discriminate the nature of observed coherences.

\section{Choosing Optimal Experimental Parameters} \label{section:optimal}

We now discuss how to use standard spectroscopic measurements to choose pulse parameters to estimate and maximize $T_W$. We conclude that linear absorption
measurements are sufficient to place reasonable bounds on $T_W$, but we will also consider the information that can be extracted from resonance-Raman and excited state spectroscopies. Appropriate choice of central frequencies will maximize the witness time, or equivalently, minimize the pulse bandwidth for which the witness will function.

In a pump-probe experiment, the experimental parameters available
are the central frequencies, pulse polarizations, pulse durations,
pulse intensities, and propagation directions. Of these, only the central
frequencies, pulse polarizations, and pulse durations impact the witness.
As long as the pulse intensity is kept within the perturbative regime,
pulse intensity only introduces an overall scaling of the signal.
Similarly, due to phase-matching considerations, the direction of
the pulse propagation affects the direction of the signal emission,
but not the actual integrated signal, as long as the pump and probe
are not collinear. The central frequencies and pulse durations are
vital to the effectiveness of the witness. Hereafter, we will assume that all the pulses share the same polarization. Setups with varying polarizations should yield qualitatively similar results.

We now present a line of reasoning for estimating
the optimal central frequency and pulse duration for a given system.
The witness functions when the emission and absorption of light become
insensitive to Stokes shifts. For this condition to hold, the pump
pulse must be sufficiently broadband to excite populations into all
excited states with significant Franck-Condon overlaps,
while the probe pulse must be sufficiently
broadband to excite population ``up'' to doubly excited states and
to stimulate emission ``down'' to the ground electronic manifold,
regardless of the initial and final nuclear configurations. Spectroscopic techniques that give information on these ``upward'' and ``downward'' vibronic transitions can thus be expected to give useful information on the optimal pulse centering and the required bandwidth. Linear absorption, resonance Raman\cite{Tannor2007}, and excited state spectroscopy \citep{Tannor2007, Boulanger1999, Miller1984} all give such information. 

We describe and justify the procedure to choose the central frequencies using analytic expressions for these spectroscopic techniques and the pump-probe signal, all within the wavepacket-overlap approach to spectroscopy \cite{Tannor2007,Yuen-Zhou2012}.  

Let us first consider the simplest case of the absorption spectrum for a system starting in a thermal distribution, where $p_{n}$ is the probability of beginning in the $n^{th}$ vibrational eigenstate of the ground state manifold, denoted by $g$. The (frequency normalized) absorption spectrum for such a system has the form \cite{Tannor2007}
\begin{align}
S_{abs}(\omega) & =\sum_{ij}\mu_{gi}\mu_{jg} \times \notag \\ 
&\sum_{\phi,n}p_{n}\braket{i,\nu_{n}^{(g)}}{\phi}\braket{\phi}{j,\nu_{n}^{(g)}}\delta(\omega_{\phi,gn}-\omega),\label{eq:S_abs}
\end{align}
where $\mu_{ab}$ is the projection of the dipole matrix element between electronic states $a$ and $b$ onto the pulse polarization, $\ket{a,\nu_{b}^{(g)}}$ is the direct product of the $a^{th}$ electronic state and the $b^{th}$ vibrational eigenstate of the ground electronic potential, $i,j$ sum over all singly excited electronic states,  $\phi$ denotes vibronic states in the singly excited manifold, and $\omega_{\alpha,\beta}$ indicates the frequency difference between states $\alpha$ and $\beta$. 

We can think of the absorption spectrum as a statistical distribution and define the mean and variance of the absorption spectrum accordingly. These concepts are useful in analyzing the necessary experimental parameters, as we can connect them to the central frequency and variance of the pulses used. We define the mean ($\bar{\omega}_{abs}$) and variance ($\Sigma_{A}^2$) of the absorption spectrum as
\begin{align}
\bar{\omega}_{abs} & =\frac{\int S_{abs}(\omega) \omega d\omega}{\int S_{abs}(\omega) d\omega},\label{eq:S_abs_mean} \\
\Sigma_{A}^2 & =\frac{\int S_{abs}(\omega) (\omega-\bar{\omega}_{abs})^2 d\omega}{\int S_{abs}(\omega) d\omega}.\label{eq:S_abs_var} 
\end{align}
We also apply this probability-distribution perspective to resonance-Raman spectroscopy in the following discussion. 

We now compare Eqs.\ \ref{eq:S_abs}--\ref{eq:S_abs_var} to the pump-probe spectrum. We follow Ref.~\citenum{Yuen-Zhou2012} and write an expression for $S_{PP}(T)=S_{SE}(T)+S_{ESA}(T)+S_{GSB}(T)$, decomposed into stimulated emission (SE), excited state absorption (ESA), and ground state bleach (GSB)  components; the full expression is presented in Appendix \ref{appendix:fullmodel}. Our analytic model assumes the dipole approximation, the Condon approximation and the rotating wave approximation with Gaussian optical pulses. We do not assume a specific number of vibrational modes nor do we include a bath other than the explicitly modeled vibrational modes.  Reference \citenum{Yuen-Zhou2012} expanded the resulting expressions to first order in powers of $\sigma_{P}$, $\sigma_{P'}$, demonstrating that the witness functions in the ultrashort-pulse limit; our goal is to choose
the central frequency of the pulses to minimize any vibrational oscillations.

We expand $S_{PP}(T)$ to second-order in $\sigma_P$, $\sigma_{P'}$ to find the most important effects of finite-duration pulses, detailed in Appendix \ref{appendix:fullmodel}. The terms in the resulting expressions can be classified as (i) non-oscillatory, (ii) oscillatory due to electronic coherences, and (iii) oscillatory due to vibrational coherences. Of these components, we are interested only in the vibrationally oscillatory (VO) components, as they can produce false positives in the witness. The vibrational oscillatory portions of the second-order terms are (see Appendix \ref{appendix:fullmodel})
\begin{widetext}
\begin{align}
 S_{SE,VO}^{(2)}(T)  &=-\frac{1}{2}\eta^4 \sigma_{P'}^{2}\sum_{ijpq}\mu_{gi}\mu_{qg}\mu_{gp}\mu_{jg}\sum_{\phi,\phi'}\sum_{n,n'}  p_{n}\braket{i,\nu_{n'}^{(g)}}{\phi}\braket{\phi}{q,\nu_{n}^{(g)}}\braket{p,\nu_{n}^{(g)}}{\phi'}\braket{\phi'}{j,\nu_{n'}^{(g)}}\nonumber \\
 & \times e^{-i(\omega_{\phi}-\omega_{\phi'})T} \left[(\omega_{\phi,gn'}-\omega_{P'})^{2}+(\omega_{\phi',gn'}-\omega_{P'})^{2}\right] \label{eq:SE2} 
 \end{align}
 \begin{align}
 S_{ESA,VO}^{(2)}(T) &=\frac{1}{2}\eta^4{\sigma_{P'}^{2}}\sum_{ijpq}\mu_{fi}\mu_{qg}\mu_{gp}\mu_{jf} \sum_{\phi,\phi'}\sum_{n,n',n'',m}  p_{n} \braket{i,\nu_{n'}^{(g)}}{\phi}\braket{\phi}{q,\nu_{n}^{(g)}}\braket{p,\nu_{n}^{(g)}}{\phi'}\braket{\phi'}{j,\nu_{n''}^{(g)}}
\nonumber \notag \\
 & \times
\braket{\nu_{n''}^{(g)}}{\nu_{m}^{(f)}}\braket{\nu_{m}^{(f)}}{\nu_{n'}^{(g)}} e^{-i(\omega_{\phi}-\omega_{\phi'})T} \left[(\omega_{fm,\phi}-\omega_{P'})^{2}+(\omega_{fm,\phi'}-\omega_{P'})^{2}\right] \label{eq:ESA2}.
 \end{align}
 %
\end{widetext}
The notation is consistent with the absorption spectrum case, with the addition that $\omega_P$, $\omega_{P'}$ are the pump and probe central frequencies, respectively, $i,j,p,q$ denote singly excited electronic states, $f$ denotes the doubly excited electronic state, and $\phi,\phi'$  denote vibronic states in the singly excited manifold. The SE and ESA terms have VO components due only to the probe pulse, while the GSB has no VO terms to second order (see Appendix \ref{appendix:fullmodel}). 

We want to choose $\omega_{P'}$ to minimize these VO terms, ideally by using information obtainable from experiments simpler than pump-probe.  Intuitively, $\omega_P$, $\omega_{P'}$ should be near the peak of the absorption spectrum, so all of the vibronic transitions are excited with approximately similar electric field amplitudes, even in the case of finite $\sigma_P$, $\sigma_{P'}$. We develop two heuristic arguments based on absorption and resonance-Raman spectroscopies to inform the choice of $\omega_P$, $\omega_{P'}$.  

Our goal is to minimize (the square integral of) the second-order VO signal given by $S^{(2)}_{VO}(T)=S^{(2)}_{SE,VO}+S^{(2)}_{ESA,VO}$.  One way to proceed is to pick a particular value of $T$ and minimize $S^{(2)}_{VO}$ at that time delay. If this $T$ gives in some sense a typical value for the VO signal, we expect such a minimization to give a result similar to the true minimization of $\int |S^{(2)}_{VO}|^2 \mathrm{d}T$.

For the stimulated emission term, consider the case $T=0$, 
\begin{align}
&S_{SE,VO}^{(2)}(0)  =\eta^4\sigma_{P'}^{2}\mu^{2}\sum_{ij}\mu_{gi}\mu_{jg}\sum_{\phi,n}p_{n} \notag \\
&\times \braket{i,\nu_{n}^{(g)}}{\phi}\braket{\phi}{j,\nu_{n}^{(g)}}(\omega_{\phi,gn}-\omega_{P'})^{2}\label{eq:SE2atT0}
\end{align}
where $\mu^2=\sum \mu_{gi}\mu_{ig}=\int S_{abs}(\omega)d\omega$. Comparison with Eqs.\ \ref{eq:S_abs} and \ref{eq:S_abs_var} shows that $S_{SE,VO}^{(2)}(0)$ is proportional to the second statistical moment of the absorption spectrum about $\omega_{P'}$. Such a signal is minimized if $\omega_{P'}$ is set to the mean of the absorption spectrum $\bar{\omega}_{abs}$, giving $S_{SE,VO}^{(2)}(0) = \eta^4\sigma_{P'}^{2}\mu^{4}\Sigma_{A}^2$.
This result provides two key pieces of information: 1) at
$T=0$, $S_{SE,VO}^{(2)}$ is minimized if the probe central frequency is the mean of the absorption spectrum and 2) ensuring the transform-limited pulse is sufficiently broadband to cover the absorption spectrum gives an indication of an upper bound on the required pulse duration. That is, if $\sigma_P < 1/\Sigma_A$, then $S_{SE,VO}^{(2)}(0) < \eta^4\mu^4 = S_{SE}^{(0)}(0)$, and we expect that the terms in the expansion of $S_{SE,VO}(\sigma,0)$ are converging. In a system with vibrational coherences only, the signal will then be dominated by the lowest order term $S_{SE,VO}^{(2)}(0)$, which monotonically decreases with decreasing pulse duration. 

We selected $T=0$ as a typical value of $S_{SE,VO}^{(2)}(T)$ because it is easily related to the absorption spectrum. The actual pump-probe signal at $T<3(\sigma_P+\sigma_{P'})$ has a pulse-overlap correction (see the short-time portion of Fig.\ \ref{PumpProbe}), but this correction does not appear in Eqs.\ \ref{eq:SE2} and \ref{eq:ESA2}. Therefore, $S_{SE,VO}^{(2)}(0)$ may be representative of typical values of $S_{SE,VO}^{(2)}(T)$. We test the validity of this heuristic argument numerically and find that when $\omega_P=\omega_{P'}=\bar\omega_{abs}$, the qualitative features of $T_W$ are reproduced by $1/\Sigma_{A}$, see Sec.\ \ref{sec:numerical}. We define a new timescale $T_A\equiv1/10\Sigma_A$ and find numerically that $T_W\geq T_A$. $T_A$ then provides a lower bound on $T_W$ that can be determined from the absorption spectrum.

A similar analysis can be performed for excited state absorption. Since the model system we use for simulations does not include a doubly excited state, however, we shall not discuss it further here, except to note that excited-state spectroscopy can be carried out in a variety of ways without necessitating ultrafast pulses or time separation between excitation and absorption. Excitation can be performed using pulsed or continuous wave lasers, or even collisional excitation. For some examples of excited state spectroscopy techniques, see Refs.~\onlinecite{Boulanger1999, Miller1984, Laming1988, Piatkowski2009, Piatkowski2008}.

The second heuristic argument originates in an examination of the wavepacket-overlap expression for a resonance-Raman experiment \citep{Tannor2007}, 
\begin{align}
&S_{R}(\omega_{S})=\int d\omega_I \sum_{p,q,i,j}\mu_{gi}\mu_{qg}\mu_{gp}\mu_{jg} \notag \\
&\times \sum_{n,n',\phi,\phi'}p_{n}\braket{i,\nu_{n'}^{(g)}}{\phi}\braket{\phi}{q,\nu_{n}^{(g)}}\braket{p,\nu_{n}^{(g)}}{\phi'}\braket{\phi'}{j,\nu_{n'}^{(g)}} \notag \\
&\times \frac{\delta(\omega_{S}-\omega_{I}+\omega_{gn',gn})}{(\omega_{I}-\omega_{\phi,gn}+i\gamma)(\omega_{I}-\omega_{\phi',gn}+i\gamma)}  ,\label{eq:ResRaman_raw}
\end{align}
where $\omega_{I}$ is the driving frequency, $\omega_{S}$ is
the emitted frequency, and $\gamma$ is the spontaneous emission rate (generally much less than $\omega_I$, $\omega_S$). In this form of resonance-Raman spectroscopy, for each $\omega_S$ we have chosen to integrate the signal over all values of $\omega_I$. Performing the integral over $\omega_I$ gives the signal 
\begin{align}
&S_{R}(\omega_{S})=\sum_{p,q,i,j}\mu_{gi}\mu_{qg}\mu_{gp}\mu_{jg} \notag \\
\times &\sum_{n,n',\phi,\phi'}p_{n}\braket{i,\nu_{n'}^{(g)}}{\phi}\braket{\phi}{q,\nu_{n}^{(g)}}\braket{p,\nu_{n}^{(g)}}{\phi'}\braket{\phi'}{j,\nu_{n'}^{(g)}} \notag \\
\times &\frac{1}{(\omega_{S}-\omega_{\phi,gn'}+i\gamma)(\omega_{S}-\omega_{\phi',gn'}+i\gamma)}.\label{eq:ResRaman}
\end{align}

We further define the resonance-Raman average frequency as 
\begin{align}
\bar{\omega}_{R} =\frac{\int S_{R}(\omega_S) \omega_S d\omega_S}{\int S_{R}(\omega_S) d\omega_S}. 
\end{align}

The resonance-Raman spectrum exhibits sharp peaks at $\omega_{S}=\omega_{\phi,gn'}$  and $\omega_{S}=\omega_{\phi',gn'}$, which  are generally smaller than the $\omega_{\phi,gn}$ dominating Eq.~\ref{eq:SE2atT0}, and has the same matrix elements and frequencies as the stimulated emission term in Eq.~\ref{eq:SE2}. We thus expect that $\bar\omega_{R}$ gives information on which intermediate states $\phi$, $\phi'$ and final vibrational state $n'$ contribute most to Eq.\ \ref{eq:SE2}. Choosing $\omega_P=\omega_{P'}=\bar\omega_R<\bar\omega_{abs}$ should then suppress these terms, suggesting that improved results might be obtained by red-shifting $\omega_P$, $\omega_{P'}$ from $\bar\omega_{abs}$ to $\bar\omega_R$.
We show numerically in Sec.\ \ref{sec:numerical} that the resonance-Raman spectrum often overestimates the optimal red shift and that, in our numerical studies, optimal results are obtained simply from the absorption spectrum. The difference between $\bar\omega_{abs}$ and $\bar\omega_R$ for some sample systems can be seen in Figs.\ \ref{WitnessBandwidth}(b) and \ref{WitnessBandwidthFC}(b).

\section{Numerical Tests of Optimal Parameters\label{sec:numerical}}

\begin{figure}
\centering 
\includegraphics[width=1\linewidth]{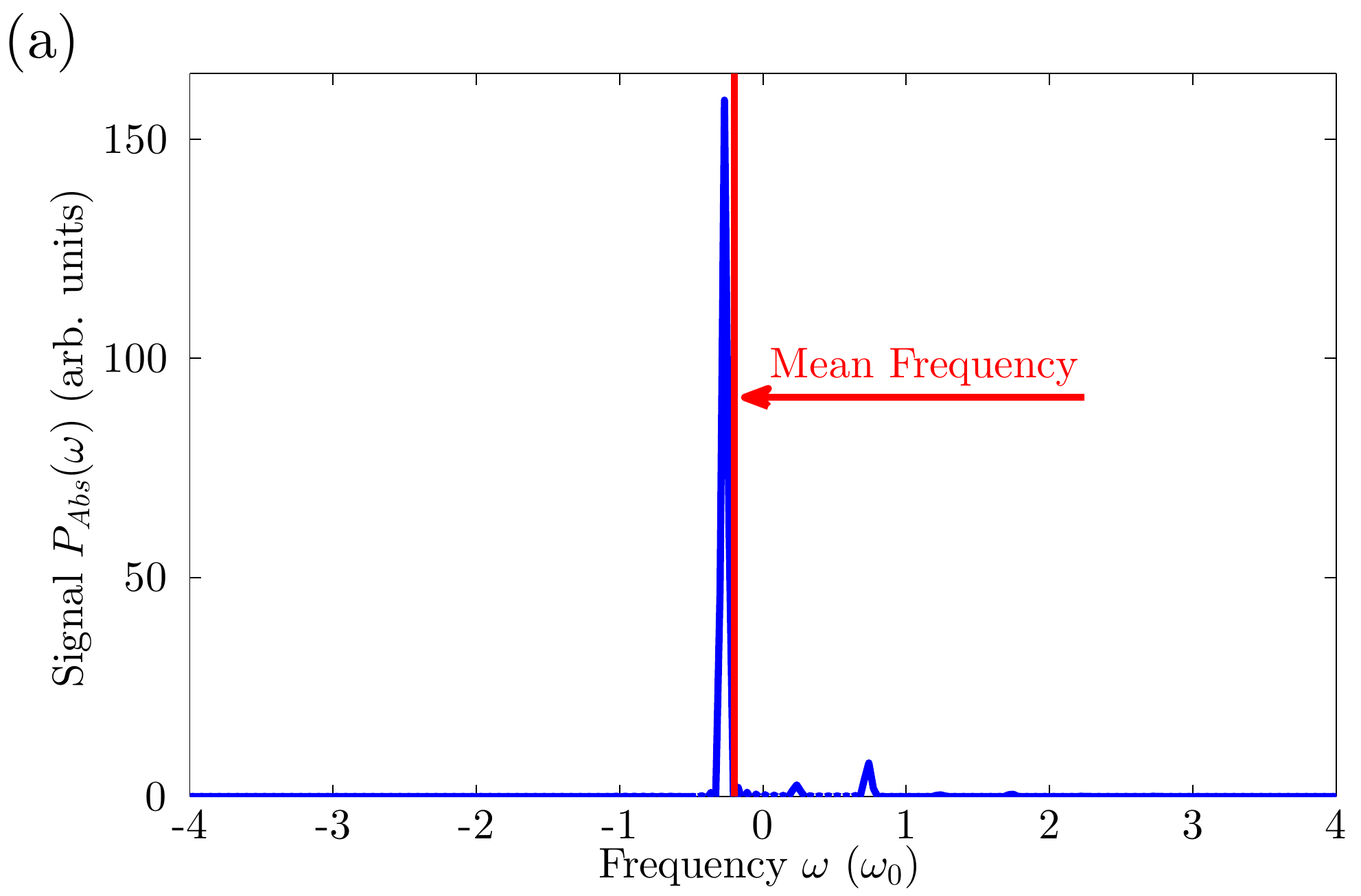}
\includegraphics[width=1\linewidth]{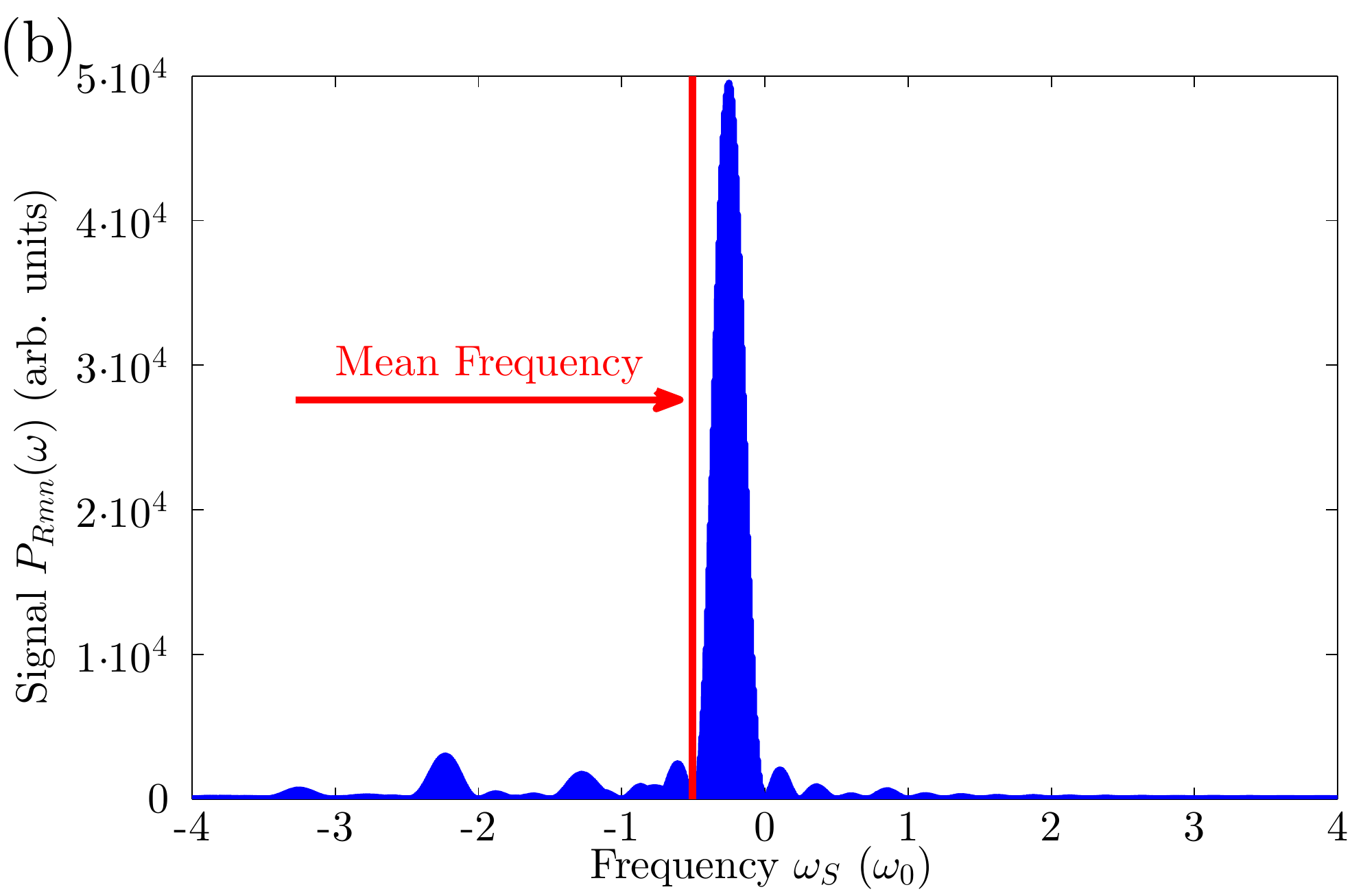}
\caption{Sample absorption (a) and Raman spectra (b) for a warped harmonic monomer with $\omega_e=0.5\omega_0$ and $\mathscr{S}=0.02$. Each spectrum results in a slightly different central frequency and gives a different variance. Frequencies are expressed with $\Omega_e$ subtracted.} \label{fig:sample_spectrums}
\end{figure}

To test the above heuristics for choosing pulse central frequencies, we performed pump-probe simulations on a monomer for a variety of parameters, with the pulses centered on $\bar\omega_{abs}$, $\bar\omega_{R}$, and $(\bar\omega_{abs}+\bar\omega_R)/2$. Example absorption and Raman spectra are shown in \ref{fig:sample_spectrums}, with the different mean frequencies indicated.

\begin{figure}[h]
\centering 
\centering 
\includegraphics[width=1\linewidth]{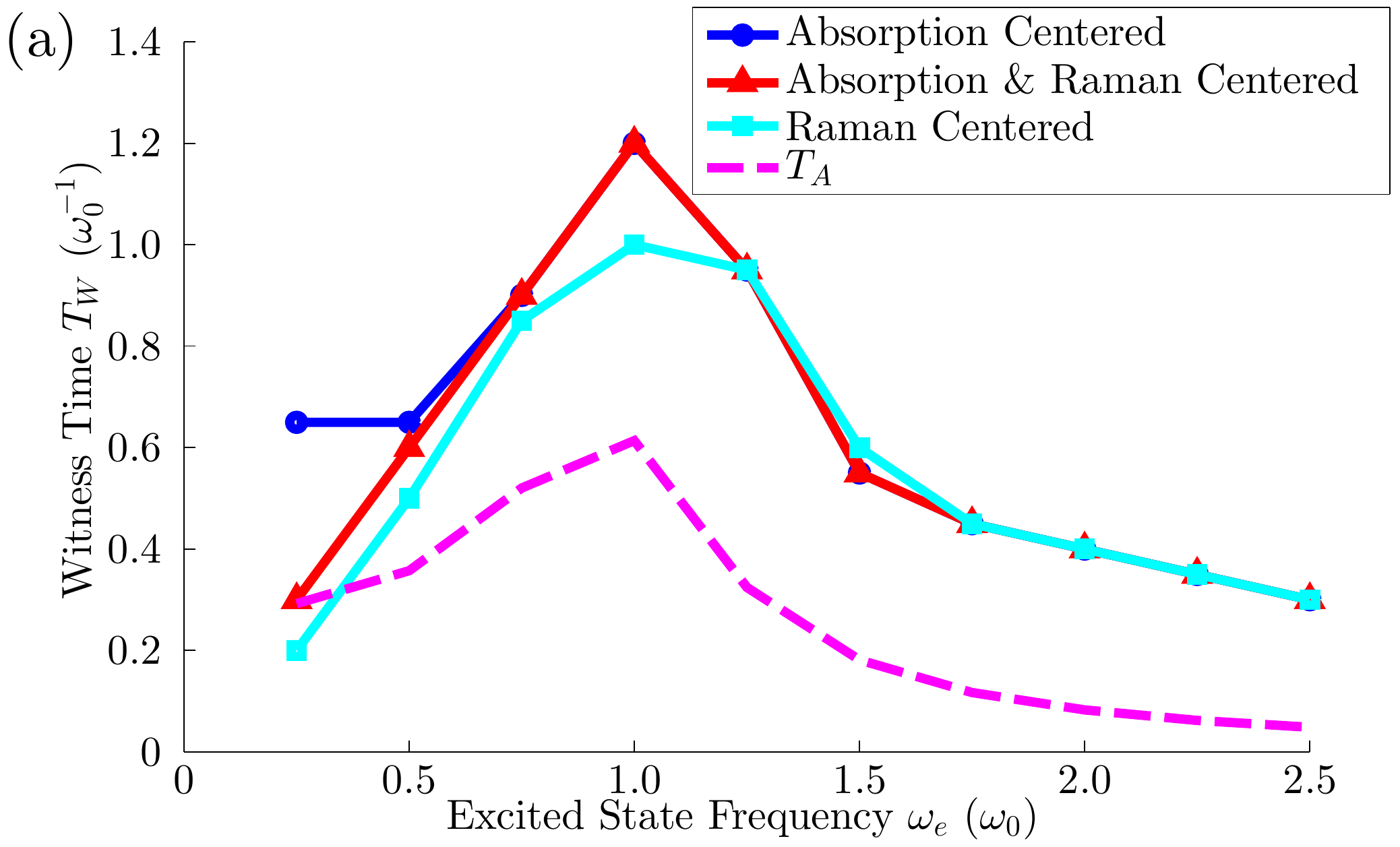}
\includegraphics[width=1\linewidth]{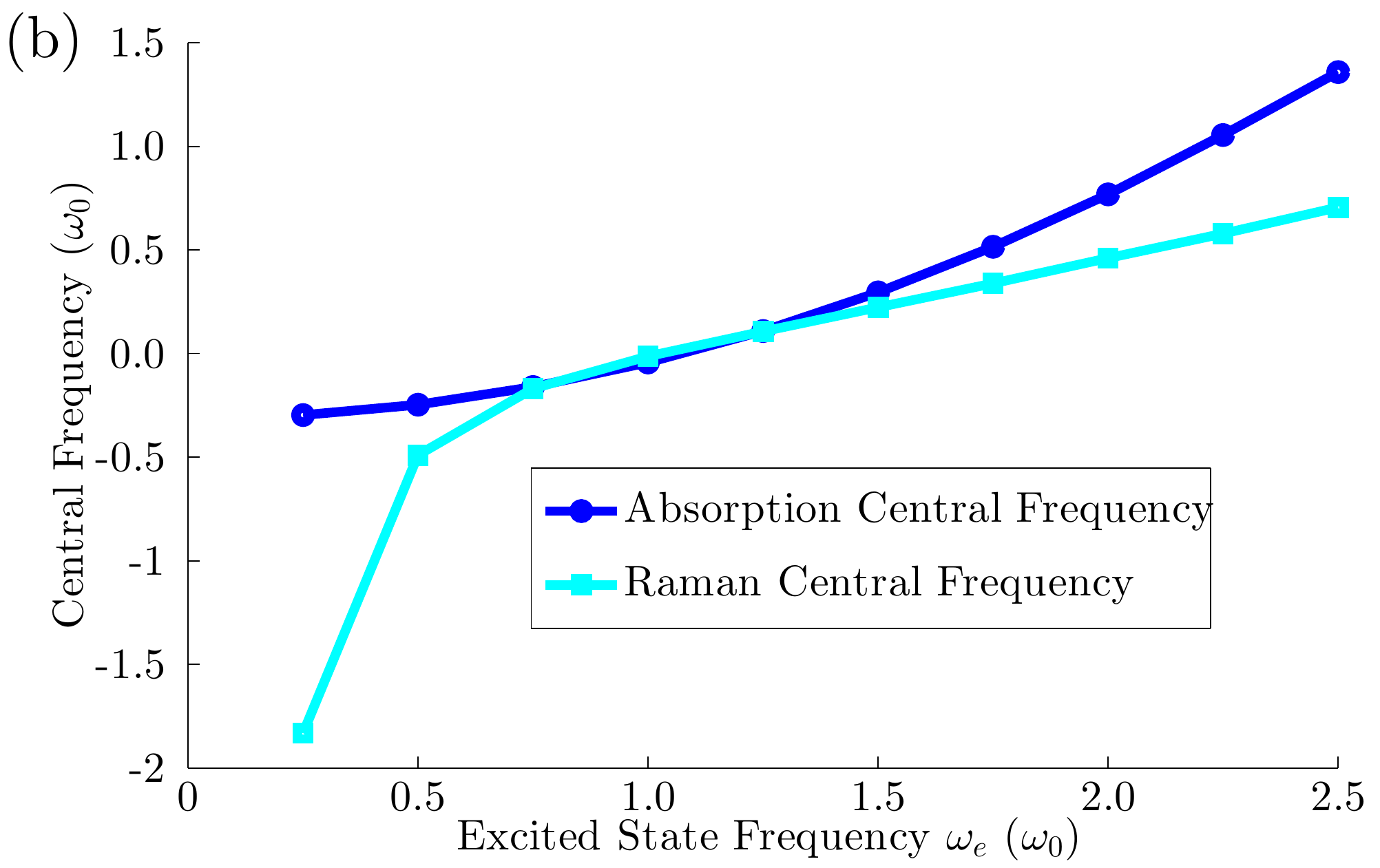}
\label{sub:frequency_scan_frequencies}

\caption{(a) Witness time vs.~excited state vibrational frequency for a monomer
with $\mathscr{S}=0.02$, using different techniques to determine the central frequencies of the pump and probe, as indicated in the legend. 
Centering using the absorption spectrum performs best over the range of excited state frequencies. The predicted witness time $T_{A}=1/10\Sigma_{A}$
captures the trend of $T_{W}$ and is less than the observed
witness time with absorption centering. (b) Central frequencies determined by the different techniques as used in a). }

\label{WitnessBandwidth} 
\end{figure}

\begin{figure}[h]
\centering
\includegraphics[width=1\linewidth]{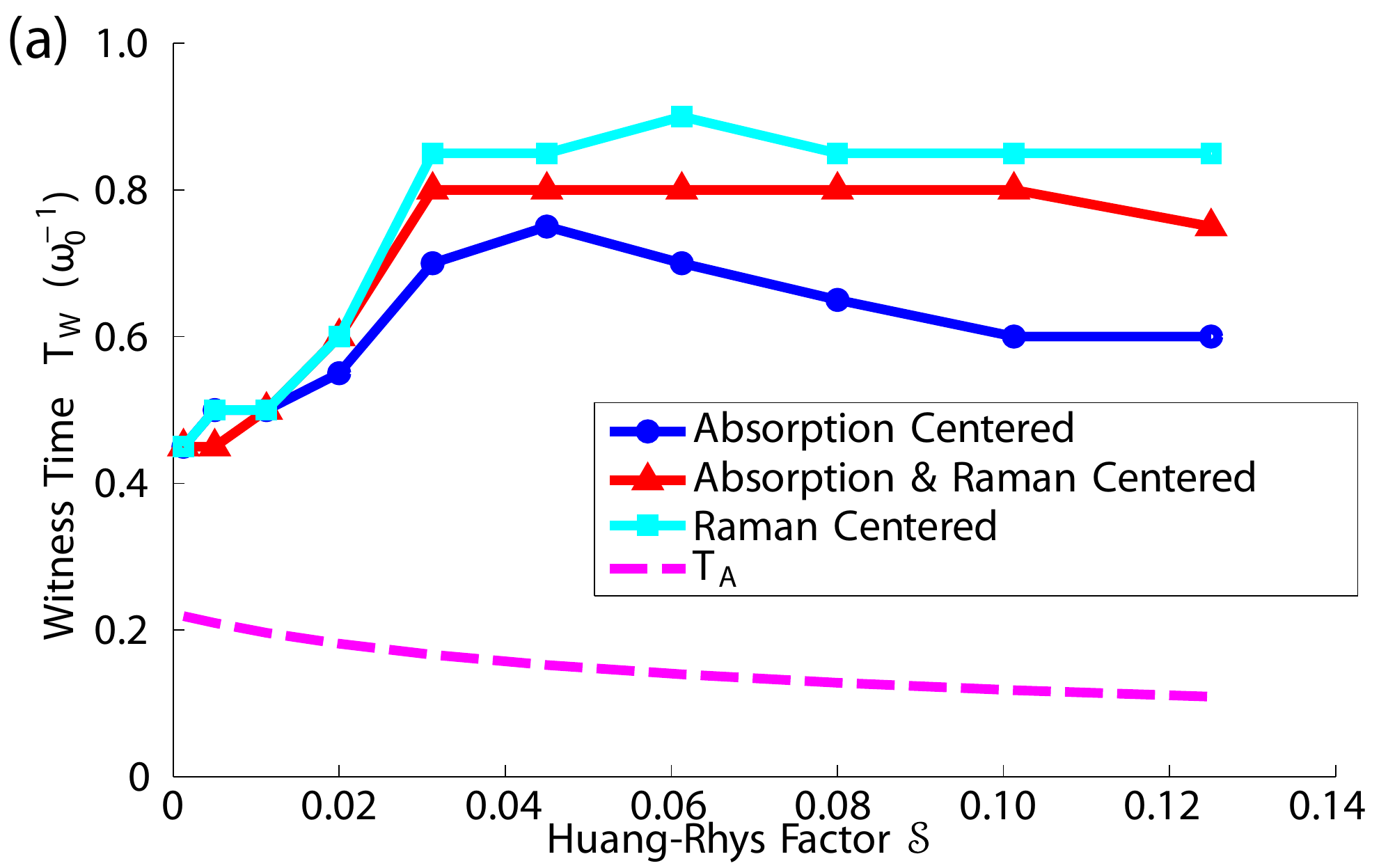}
\includegraphics[width=1\linewidth]{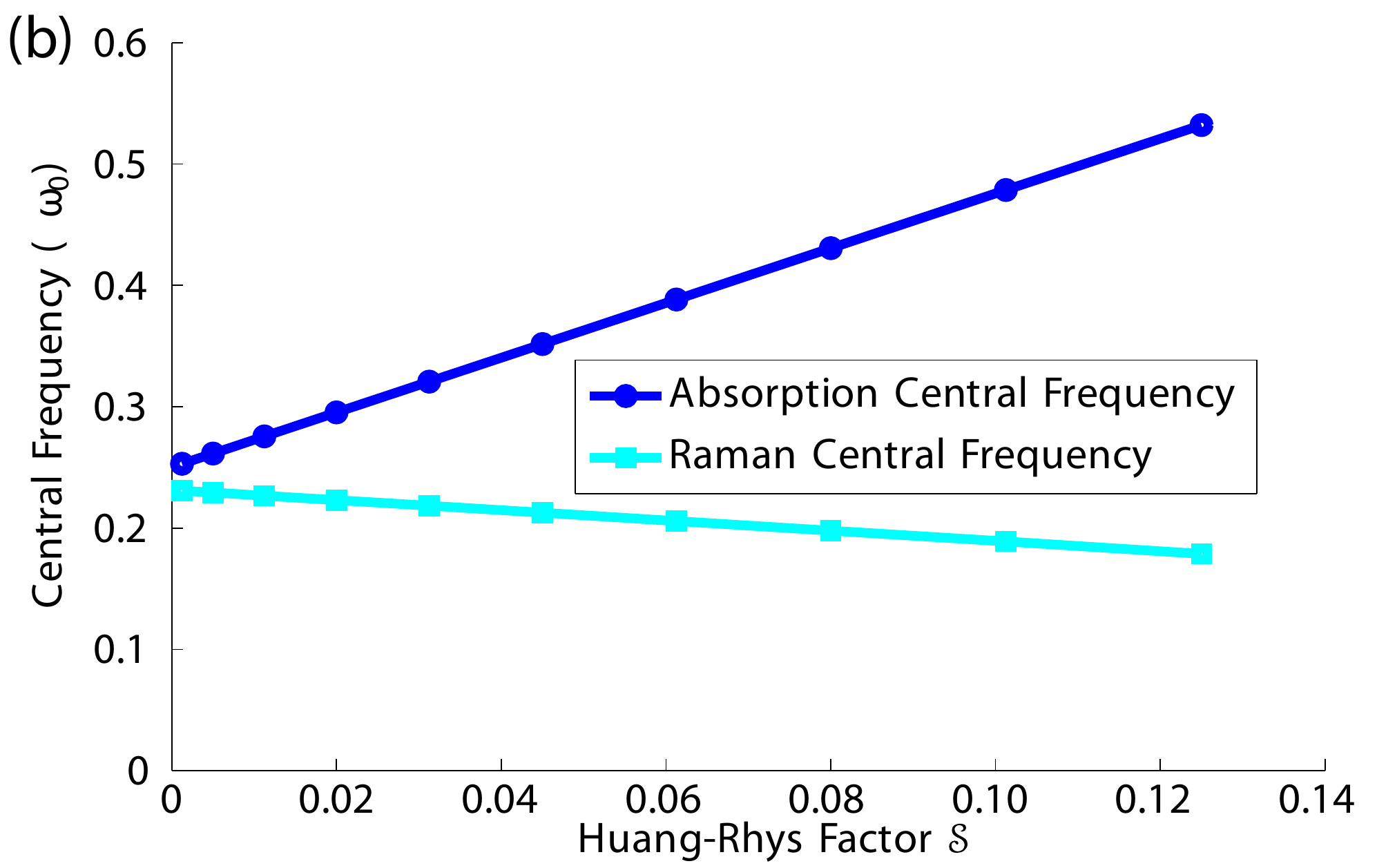}
\caption{(a) Witness time vs.\ Huang-Rhys factor for different centering techniques, as in Fig.~\ref{WitnessBandwidth}, with excited state vibrational frequency $\omega_e=1.5\omega_0$. As in Fig.\ref{WitnessBandwidth}, $T_A$ is less than the observed witness time with absorption centering. (b) Central frequencies determined by the different techniques, as used in (a).}

\label{WitnessBandwidthFC} 
\end{figure}

Figures \ref{WitnessBandwidth} and \ref{WitnessBandwidthFC} show the witness time for a variety of excited state vibrational frequencies and Huang-Rhys factors. Pulse central frequencies chosen from the three spectral methods are shown.  Centering using the absorption spectrum
generally maximizes the witness time, while centering using the resonance-Raman
spectra gives poor results with low-frequency excited state vibrations. At large Huang-Rhys factors, the Raman spectrum centering method performs best.
Over a large range of parameters, the three methods give similar results, suggesting that while the witness is sensitive to the correct choice of central
frequency, it is robust to small variations about the optimal value.

Also plotted in Figs.~\ref{WitnessBandwidth} and \ref{WitnessBandwidthFC} is the time scale $T_{A}=1/10\Sigma_{A}$. Both figures show that $T_A$ accurately
captures the dependence of the absorption-centered witness time, and
by using pulses with a duration $\sigma_P=\sigma_{P'}<T_{A}$ the witness
is always seen to function. In Fig.\ \ref{WitnessBandwidthFC}, $T_{A}$ forms an aggressive lower bound for
the witness time $T_W$ as $\mathscr{S}$ increases, showing that the witness may be implementable with considerably longer pulses. 

Based on these results,
we propose a prescription for determining optimal experimental parameters for
the witness: obtain the absorption spectrum, center the pump
and probe pulses on its mean frequency $\bar\omega_{abs}$, and ensure the pulses have
a time duration variance of less than $T_{A}^2$ (or FWHM of less than $2\sqrt{2\ln2}T_{A}$). Alternatively, estimates
of the required pulse durations can be obtained through simulations
or comparisons to model systems.

\begin{figure}[h]
\centering \includegraphics[width=\figwidth]{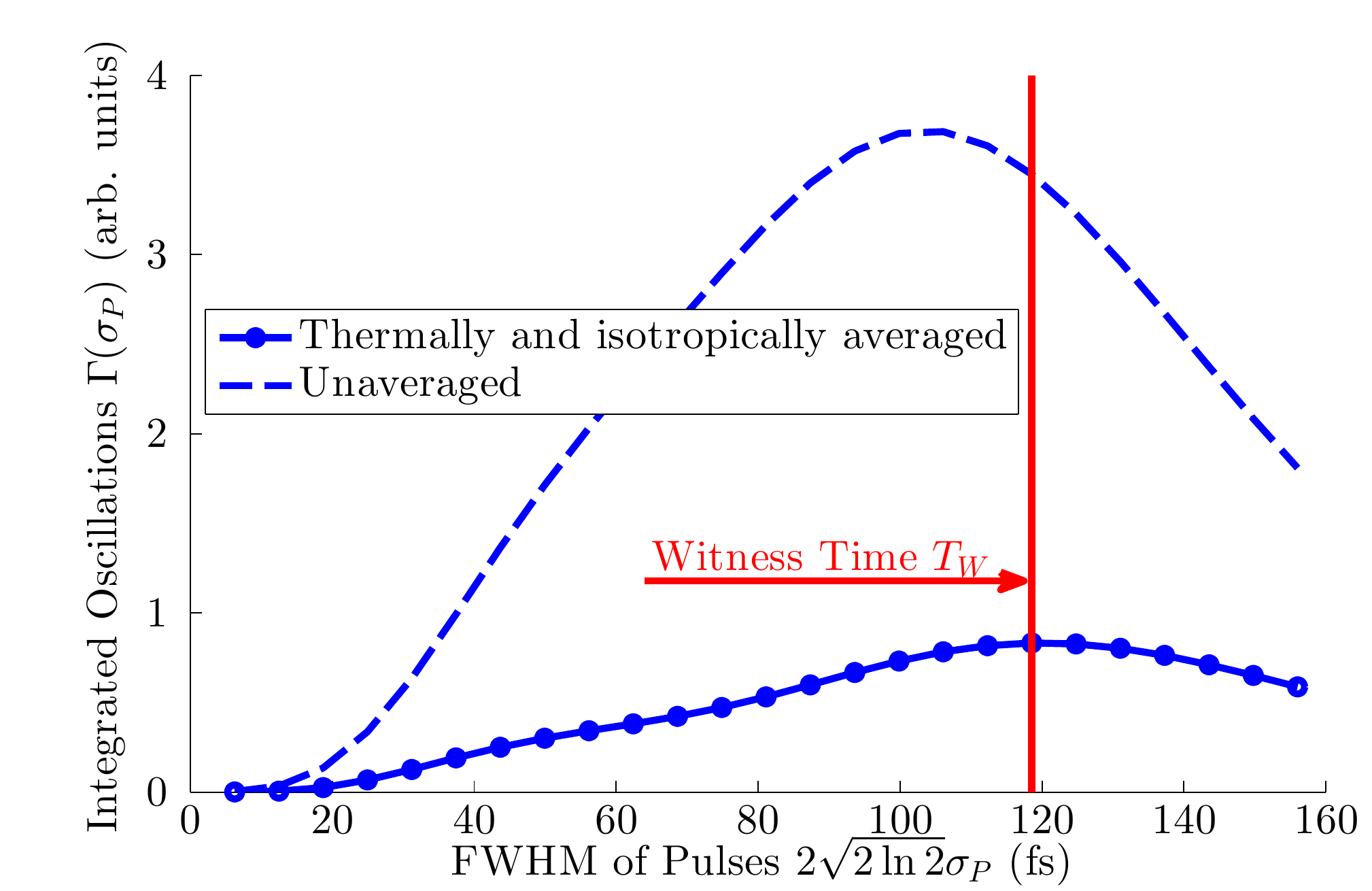}

\caption{Witness plot for a monomer at 294~K, including full thermal and isotropic
averaging. The ground state frequency is $100$~cm$^{-1}$, and the
witness time is $118$~fs, with $\omega_{e}=1.5\omega_0$ and $\mathscr{S}=0.02$. The pulse duration is given in FWHM to better compare with experimental measures. Also plotted is the witness curve resulting from a system originally only populating the ground state. The averaging shifts $T_W$ from 106 fs in the ground state system to 118 fs in the full system.}

\label{Witness_thermal} 
\end{figure}

In order to apply our results to physical systems, we fix $\omega_{0}=100$~cm$^{-1}$, which is similar to important vibrational modes in a prototypical photosynthetic complex, the Fenna-Matthews-Olson complex \cite{Adolphs2008}. This choice gives a time unit of $53$~fs. Fig.\ \ref{Witness_thermal} shows results similar to Fig.~\ref{Witness_monomer} but with full thermal and isotropic averaging at temperature 294~K. The system has the same parameters as considered in Figures \ref{PumpProbe} and \ref{Witness_monomer}, with an excited state vibrational frequency of $\omega_{e}=1.5\omega_0$ and $\mathscr{S}=0.02\omega_0^{-1}$. The pump and probe pulses are centered at $\bar\omega_{abs}$.  For these parameters, representative of a broad variety of real physical systems, pulses of FWHM $\approx120$~fs or shorter are sufficient to perform the witness. Pulses with FWHM in this regime are routinely accessible with modern mode-locked laser systems. This suggests the technique could be implemented in many ultrafast spectroscopy labs.

\section{Summary and conclusion}

We have developed a practical implementation of a recently proposed
witness for electronic coherences, in which   oscillations in the frequency-integrated pump-probe signal are uniquely attributable to electronic coherences for optical pulses in the impulsive limit. By performing pump-probe experiments at a variety of pulse durations and extrapolating to the impulsive limit, vibrational and electronic coherences can be discriminated by examining if the oscillatory signals are monotonically decreasing or increasing. This witness functions only for pulses with a duration less than the witness time $T_{W}$. We have further shown both analytically and numerically that $T_{W}$ can be maximized by centering the pump and probe pulses on the mean of the absorption spectrum. The witness time is found numerically to have a lower bound given by $1/10\Sigma_{A}$, in terms of the variance of the absorption spectrum $\Sigma^2_{A}$.  For parameters chosen for a model system of biological relevance, pulses with FWHM of approximately 120~fs are sufficient for discriminating between electronic and vibrational coherences. For systems with faster relevant vibrations, the requirements for pulse durations become accordingly stricter. This witness technique is thus readily implementable with commercially available laser systems, and can potentially resolve the debate over the nature of coherences observed in photosynthetic complexes, as well as have wider applications in examining coherent processes in physical systems.

\section*{Acknowledgments}

This work was enabled by the use of computing resources provided by Odyssey, supported by the FAS Science Division Research Computing Group at Harvard University, WestGrid, and Compute/Calcul Canada. J.\ J.\ K. and A.\ S.\ J.\ acknowledge support from NSERC. A.\ S.\ J. acknowledges support from the European Union Marie Curie program. A.\ A.-G.\ and J.\ Y.-Z.\ acknowledge support from the Center for Excitonics, an Energy Frontier Research Center funded by the U.S. Department of Energy, Office of Science and Office of Basic Energy Sciences, under Award Number DE-SC0001088. 

\appendix
\section{Numerical Simulations}
\label{appendix:numerics}
The model used in our numerical simulations is outlined in Section \ref{section:background}; here we will discuss the extension from a monomer to the case of a coherent dimer and describe the numerical methods. 

In extending the model to a dimer, we assume each of the two sites has a single electronic excitation and vibrational mode. Each site's vibrational mode is treated as a harmonic potential surface for the generalized vibrational coordinate. Both ground state vibrational modes have frequency $\omega_0$. Dipole-allowed transitions are from the ground to singly excited states, and from the singly excited to doubly excited state. We assume the doubly excited state does not exhibit any binding energy, and the transition dipoles are oriented $90^\circ$ apart with the ratio of their norms $1:3$. The full dimer Hamiltonian can then be written as

\begin{widetext}
\begin{align}
H&= \frac{p_1^2+p_2^2}{2} +\frac{\omega_0^2(x_1^2+x_2^2)}{2}|g\rangle\langle g|+\big( \Omega_{e,1}+\frac{\omega_{e,1}^2(x_1-\Delta_{x_1})^2 + \omega_0^2x_2^2}{2}\big)|e_1\rangle\langle e_1|
\notag \\ 
&+\big(\Omega_{e,2}+\frac{\omega_0^2x_1^2+\omega_{e,2}^2(x_2-\Delta_{x_2})^2}{2}\big)|e_2\rangle\langle e_2|+\big(\Omega_{e,1}+\Omega_{e,2}+\frac{\omega_{e,1}^2(x_1-\Delta_{x_1})^2+\omega_{e,2}^2(x_2-\Delta_{x_2})^2}{2}\big)|f\rangle\langle f| \notag \\
&+ J(|e_1\rangle\langle e_2|+|e_2\rangle\langle e_1|) \label{eq:dimer_hamil}
\end{align}
\end{widetext}
where $x_i$  (for $i=1,2$) is the nuclear coordinate including the particle masses, $p_i$ is the conjugate momentum, $|e_i\rangle$ denotes the electronic state, $\omega_{e,i}$ is the excited state vibrational frequency, $\Omega_{e,i}$ is the electronic excitation energy, $\Delta_{x,i}$ is the equilibrium vibrational coordinate in the electronically excited state, $J$ is the coupling between the two singly excited states, and Planck's constant $\hbar$ is set to 1.

In order to perform the numerical simulations, we treat the light-matter interaction perturbatively and define perturbative higher-order wavefunctions as in References \citenum{Yuen-Zhou2014}, \citenum{Yuen-Zhou2012}, and \citenum{ Tannor2007}. These wavefunctions permit calculation of the pump-probe, absorption, and resonance-Raman spectra. We propagate these wavefunctions using the split-operator method \citep{Leforestier1991}, treating the action of the pulses as a perturbation moving the vibrational wavepacket from one electronic state to another. Full details of our numerical method can be found in Reference \citenum{Yuen-Zhou2014}. The vibrational wavepackets were simulated on a grid of 30 points per mode, with a spacing of $0.5\omega_0^{-0.5}$. The total time simulated was $25\omega_0^{-1}$, in steps of $0.01\omega_0^{-1}$, with the action of the pulses treated at all times. These parameters were rigorously tested for convergence. In order to avoid numerical errors we used a complex absorbing potential barrier. The complex barrier was modelled by the Eckart potential \citep{Kosloff1986}, with a width of $3\omega_0^{-0.5}$ and a complex amplitude of $-10(1+i)\omega_0$. As the wavefunctions of a harmonic oscillator are spatially confined, this absorbing barrier was not generally sampled, but was included as a precaution and for extensions to non-harmonic potentials.

\section{Full Model Expressions}
\label{appendix:fullmodel}
In the main text we use a wavepacket description of the pump-probe signal in order to derive bounds on the required pulse durations for the witness to function. In Sec.\ \ref{section:optimal}, these bounds were found by expanding the full model expressions in powers of the pump and probe pulse durations, and considering the pulse durations for which this expansion can reasonably be expected to converge. In this appendix we derive the expansion up to second order, and demonstrate that the terms presented in equation \ref{eq:SE2} are the only vibrationally oscillatory terms. 

We use the model of Ref.\ \citenum{Yuen-Zhou2012}. Unlike in the numerics presented, we do not assume either a monomer or a dimer; instead, the system considered analytically is  more general. In particular, we assume neither a number of electronic nor vibrational modes -- in essence, we allow for an unlimited number of coupled, singly excited electronic states, though for convenience we assume a single doubly excited state and that the ground-to-doubly-excited-state transition is dipole forbidden. 

In equations B9-B11 of reference \citenum{Yuen-Zhou2012}, it was shown that the pump-probe spectrum can be written as the sum of three terms, $S_{PP}(T)= S_{GSB}(T) + S_{SE}(T) + S_{ESA}(T)$, with those terms given by: 
\begin{widetext}
\begin{align}
S_{SE}(T)&=\sum_{ijpq}\mu_{gi}\mu_{qg}\mu_{gp}\mu_{jg} \sum_{\phi,\phi',n,n'} p_n  \braket{i,\nu_{n'}^{(g)}}{\phi}\braket{\phi}{q,\nu_n^{(g)}} \braket{p,\nu_{n}^{(g)}}{\phi'}\braket{\phi'}{j,\nu_{n'}^{(g)}} e^{-i(\omega_{\phi}-\omega_{\phi'})T} \label{eq:SE_full}\\ \notag
&\times \epsilon_{P'}(\omega_{\phi,gn'})\epsilon_{P}(\omega_{\phi,gn})\epsilon_{P}(\omega_{\phi',gn})\epsilon_{P'}(\omega_{\phi',gn})  \\ 
S_{ESA}(T)&=-\sum_{ijpq}\mu_{fi}\mu_{qg}\mu_{gp}\mu_{jf} \sum_{\phi,\phi',n,n',n'',m} p_n \braket{i,\nu_{n'}^{(g)}}{\phi}\braket{\phi}{q,\nu_n^{(g)}} \braket{p,\nu_{n}^{(g)}}{\phi'}\braket{\phi'}{j,\nu_{n''}^{(g)}}\braket{\nu_{n''}^{(g)}}{\nu_{m}^{(f)}}\braket{\nu_{m}^{(f)}}{\nu_{n'}^{(g)}}  \label{eq:ESA_full} \\ 
&\times \epsilon_{P'}(\omega_{fm,\phi})\epsilon_{P}(\omega_{\phi,gn})\epsilon_{P}(\omega_{\phi',gn})\epsilon_{P'}(\omega_{fm,\phi'})e^{-i(\omega_{\phi}-\omega_{\phi'})T}\notag\\
S_{GSB}(T)&=\Re \sum_{ijpq}\mu_{gi}\mu_{qg}\mu_{gp}\mu_{jg} \sum_{\phi,\phi',n,n'} p_n \epsilon_{P'}(\omega_{\phi,gn})\epsilon_{P'}(\omega_{\phi,gn'})\epsilon_{P}(\omega_{\phi',gn})\epsilon_{P}(\omega_{\phi',gn'}) \times  \label{eq:GSB_full} \\ \notag
& \braket{i,\nu_{n}^{(g)}}{\phi}\braket{\phi}{q,\nu_{n'}^{(g)}} \braket{p,\nu_{n'}^{(g)}}{\phi'}\braket{\phi'}{j,\nu_{n}^{(g)}} e^{-i \omega_{gn',gn} T} 
 \left[1-\text{Erf}\left(-i\sigma_P\frac{2\omega_P-\omega_{\phi',gn'}-\omega_{\phi',gn}}{2}\right)\right] \\ \notag
\end{align}
\end{widetext}
where we have corrected typos in equation B11 from Ref.\ \citenum{Yuen-Zhou2012} related to the complex conjugations, as well as typos in the signs of the frequencies, and have relabelled the variable $\xi$ to $\phi$. Furthermore we have corrected the expression for GSB extensively, starting from the first line of equation A12 from Ref.\ \citenum{Yuen-Zhou2012}. Here $\epsilon_{Q}$ denotes the electric field amplitude of each pulse at frequency $\omega$, given by $\epsilon_{Q}(\omega)=\eta e^{-\sigma_{Q}^{2}(\omega-\omega_{Q})^{2}/2}$, where $Q=P,P^{\prime}$ for the pump and probe pulses, respectively.  When $\sigma_{P,P'}\rightarrow 0$, $S_{GSB}(T)$ is a constant and $S_{SE}$ and $S_{ESA}$ only oscillate if there are coupled electronic states \cite{Yuen-Zhou2012}. To study the effects of finite-duration pulses, we can expand equations \ref{eq:SE_full}-\ref{eq:GSB_full} in powers of $\sigma_P$ and $\sigma_{P'}$. 

Expanding these expressions in terms of $\sigma_P$ and $\sigma_{P'}$, the first order terms of $S_{SE}$ and $S_{ESA}$ vanish due to the symmetric Gaussian profile of the pulses, as shown in Ref.\ \citenum{Yuen-Zhou2012}. The remaining first order term is from the ground state bleach, arising from the pulse-overlap error functions. 
\begin{align}
&S_{GSB}^{(1)}(T)=\Re  \eta^4\sum_{ijpq}\mu_{gi}\mu_{qg}\mu_{gp}\mu_{jg} \sum_{\phi,\phi',n,n'} p_n \times \notag \\ 
& \braket{i,\nu_{n}^{(g)}}{\phi}\braket{\phi}{q,\nu_{n'}^{(g)}} \braket{p,\nu_{n'}^{(g)}}{\phi'}\braket{\phi'}{j,\nu_{n}^{(g)}} e^{-i \omega_{gn',gn} T}  \\
& \times \left(-i\sigma_P\frac{2\omega_P-\omega_{\phi',gn'}-\omega_{\phi',gn}}{\sqrt{\pi}}\right) \notag 
\end{align}
The sum no longer depends on $\phi$. Performing a sum over $\phi$ yields $\braket{\nu_n^{(g)}}{\nu_{n'}^{(g)}}=\delta_{n,n'}$. The phase term then vanishes, and the ground state bleach becomes
\begin{align*}
&S_{GSB}^{(1)}(T)=2\Im \eta^4 \sigma_{P} \sum_{ipq}\mu_{gi}\mu_{ig}\mu_{gj}\mu_{jg} \notag \\
&\times \sum_{\phi,n} p_n \braket{i,\nu_{n}^{(g)}}{\phi'}\braket{\phi'}{j,\nu_n^{(g)}}  \frac{(\omega_{\phi',gn}-\omega_{P})}{\sqrt{\pi}}.
\end{align*}
To first order then, the ground state bleach also yields a constant background. 

The next order terms are second order in $\sigma_P, \sigma_{P'}$, arising from the Gaussian pulse profiles and cross terms in the pulse-overlap terms. These second order terms are:
\begin{widetext}
\begin{align*}
S_{SE}^{(2)}(T)&=-\frac{1}{2}\eta^4 \sum_{ijpq}\mu_{gi}\mu_{qg}\mu_{gp}\mu_{jg} \sum_{\phi,\phi',n,n'} p_n  \braket{i,\nu_{n'}^{(g)}}{\phi}\braket{\phi}{q,\nu_n^{(g)}} \braket{p,\nu_{n}^{(g)}}{\phi'}\braket{\phi'}{q,\nu_{n'}^{(g)}} e^{-i(\omega_{\phi}-\omega_{\phi'})T} \\ \notag
&\times \left[\sigma_{P'}^2(\omega_{\phi,gn'}-\omega_{P'})^2+{\sigma_{P'}^2}{(\omega_{\phi',gn'}-\omega_{P'})^2} + {\sigma_P^2}{(\omega_{\phi,gn}-\omega_P)^2}+{\sigma_P^2}{(\omega_{\phi',gn}-\omega_P)^2} \right]\\
S_{ESA}^{(2)}(T)&=\frac{1}{2}\eta^4\sum_{ijpq}\mu_{gi}\mu_{qg}\mu_{gp}\mu_{jg} \sum_{\phi,\phi',n,n',n'',m} p_n \braket{i,\nu_{n'}^{(g)}}{\phi}\braket{\phi}{q,\nu_n^{(g)}} \braket{p,\nu_{n}^{(g)}}{\phi'}\braket{\phi'}{q,\nu_{n''}^{(g)}}\braket{\nu_{n''}^{(g)}}{\nu_{m}^{(f)}}\braket{\nu_{m}^{(f)}}{\nu_{n'}^{(g)}} e^{-i(\omega_{\phi}-\omega_{\phi'})T} \\ \notag
&\times \left[{\sigma_{P'}^2}{(\omega_{fm,\phi}-\omega_{P'})^2}+{\sigma_{P'}^2}{(\omega_{fm,\phi'}-\omega_{P'})^2} + {\sigma_P^2}{(\omega_{\phi,gn}-\omega_P)^2}+{\sigma_P^2}{(\omega_{\phi',gn}-\omega_P)^2}\right]\\
S_{GSB}^{(2)}(T)&=-\frac{1}{2} \eta^4 \Re \sum_{ijpq}\mu_{gi}\mu_{qg}\mu_{gp}\mu_{jg} \sum_{\phi,\phi',n,n'} p_n   \braket{i,\nu_{n}^{(g)}}{\phi}\braket{\phi}{q,\nu_{n'}^{(g)}} \braket{p,\nu_{n'}^{(g)}}{\phi'}\braket{\phi'}{j,\nu_{n}^{(g)}}e^{-i \omega_{gn',gn} T} \\ \notag
&\times \left[\sigma_{P'}^2(\omega_{\phi,gn}-\omega_{P'})^2+\sigma_{P'}^2(\omega_{\phi,gn'}-\sigma_{P'})^2+\sigma_{P}^2(\omega_{\phi',gn}-\omega_P)^2+ \sigma_{P}^2(\omega_{\phi',gn'}-\omega_P)^2\right]
\\ \notag
\end{align*}
\end{widetext}

In the ground state bleach contribution, the pulse-overlap terms once again admit a sum over either $\phi$ or $\phi'$, giving $\braket{\nu_n^{(g)}}{\nu_{n'}^{(g)}}=\delta_{n,n'}$, removing any oscillatory contributions. This yields
\begin{align*}
S_{GSB}^{(2)}(T)&=-\eta^4 \Re \mu^2 \sum_{jp}\mu_{gj}\mu_{pg}\sum_{\phi,n} p_n   \braket{j,\nu_{n}^{(g)}}{\phi}\braket{\phi}{p,\nu_{n'}^{(g)}} \\ \notag
&\times \left[\sigma_{P'}^2(\omega_{\phi,gn}-\omega_{P'})^2+\sigma_{P}^2(\omega_{\phi',gn}-\omega_P)^2\right]
\\ \notag
\end{align*}
which is again a non-oscillatory contribution.

The terms dependent on $\sigma_P$ in the ESA and SE contributions have no dependence on $n'$, and thus admit a summation over the nuclear degree of freedom. Any oscillations in this term are then due to the electronic degree of freedom, and do not affect the witness. The surviving vibrational oscillatory (VO) terms are then given by 

\begin{widetext}
\begin{align}
 S_{SE,VO}^{(2)}(T)  &=-\frac{1}{2}\eta^4 \sigma_{P'}^{2}\sum_{ijpq}\mu_{gi}\mu_{qg}\mu_{gp}\mu_{jg}\sum_{\phi,\phi'}\sum_{n,n'}^{\infty}  p_{n}\braket{i,\nu_{n'}^{(g)}}{\phi}\braket{\phi}{q,\nu_{n}^{(g)}}\braket{p,\nu_{n}^{(g)}}{\phi'}\braket{\phi'}{j,\nu_{n'}^{(g)}}\nonumber \\
 & \times e^{-i(\omega_{\phi}-\omega_{\phi'})T} \left[(\omega_{\phi,gn'}-\omega_{P'})^{2}+(\omega_{\phi',gn'}-\omega_{P'})^{2}\right] \notag \\
 S_{ESA,VO}^{(2)}(T) &=\frac{1}{2}\eta^4{\sigma_{P'}^{2}}\sum_{ijpq}\mu_{fi}\mu_{qg}\mu_{gp}\mu_{jf} \sum_{\phi,\phi'}\sum_{n,n',n'',m}^{\infty}  p_{n} \braket{i,\nu_{n'}^{(g)}}{\phi}\braket{\phi}{q,\nu_{n}^{(g)}}\braket{p,\nu_{n}^{(g)}}{\phi'}\braket{\phi'}{j,\nu_{n''}^{(g)}}
\nonumber \notag \\
 & \times
\braket{\nu_{n''}^{(g)}}{\nu_{m}^{(f)}}\braket{\nu_{m}^{(f)}}{\nu_{n'}^{(g)}} e^{-i(\omega_{\phi}-\omega_{\phi'})T} \left[(\omega_{fm,\phi}-\omega_{P'})^{2}+(\omega_{fm,\phi'}-\omega_{P'})^{2}\right], \notag \\
\end{align}
\end{widetext}
which are Eqs.\ \ref{eq:SE2} and \ref{eq:ESA2}. We conclude that for the witness to function, the SE and ESA contributions require $\sigma_{P'}$ to be small (i.e., a short probe pulse) while the GSB contributions do not affect the witness to second order in pulse durations. These model expressions are compared to other spectroscopic quantities in the main text.

%

\end{document}